\documentclass[twocolumn,showpacs,prl,footinbib,a4paper,superscriptaddress,floatfix]{revtex4}

\usepackage{amssymb,amsmath}
\usepackage{graphics}
\usepackage{dcolumn}
\usepackage{bm}
\usepackage{epsfig}

\newcommand{\expect}[1]{\langle #1 \rangle}
\newcommand{\GF}{{\rm Green's function}\ }
\newcommand{\GFs}{{\rm Green's functions}\ }

\newcommand{\NE}{{\rm non-equilibrium}\ }

\newcommand{\tr}{{\rm Tr}}
\usepackage[usenames,dvipsnames]{color}
\definecolor{MyOrange}{rgb}{1.0,0.5,0}
\definecolor{MyPurple}{rgb}{0.5,0,1}

\begin{document}


\title{Non-equilibrium charge susceptibility and dynamical conductance:
Identification of scattering processes in quantum transport}

\author{H. Ness}
\email{herve.ness@york.ac.uk}
\author{L. K. Dash}

\affiliation{Department of Physics, University of York, Heslington, York YO10 5DD,
UK}
\affiliation{European Theoretical Spectroscopy Facility (ETSF)}

\date{\today}

\begin{abstract}
  We calculate the \NE charge transport properties of nanoscale junctions in the steady
  state and 
  extend the concept of charge susceptibility to the \NE conditions. 
  We show that the \NE charge susceptibility is related to the non-linear
  dynamical conductance. 
  In spectroscopic terms, both contain the same features versus applied bias when charge 
  fluctuation occurs in the corresponding electronic resonances. 
  However, we show that while the conductance exhibits features at biases corresponding
  to inelastic scattering with no charge fluctuations, the \NE charge susceptibility does
  not. 
  We suggest 
  that measuring both the \NE conductance and charge susceptibility in the same experiment will
  permit us to differentiate between different scattering processes in quantum transport.
\end{abstract}
 

\maketitle

Recent developments in modern techniques of microscopic manipulation and nanotechnologies
enable us to build functional nanoscale systems, for example, electronic nanodevices or 
molecular motors \cite{Ratner:Book2002,Reed:Book2003,Cuniberti:Book2005,DiVentra:Book2008,
Joachim:Book2009}. 
In such systems, classical equilibrium thermodynamics is not suited to describe the quantization 
of the charge or heat flow. 
The properties of such devices differ from their equilibrium counter parts since \NE quantum and 
non-linear effects dominate.
The concepts of conventional statistical mechanics and linear-response theory for these 
small systems need to be substituted with those of non-equilibrium quantum statistical 
mechanics \cite{McLennan:1959,Zubarev:1994,Hershfield:1993}. This is
the appropriate framework for dealing with nanoscale systems driven out of equilibrium,
especially when one wants to design or control these systems as heat engines or 
electro-mechanical devices.

Extending the concept of equilibrium statistical mechanics (linear-response theory and
response functions, fluctuation theorems) to the non-equilibrium (NE) conditions has seen a
recent growing interest \cite{Esposito:2009,Shimizu:2010,Campisi:2011}.
It is known from linear-response theory that there exists some relationship between different response
functions, like for example, the density-density, current-density, or current-current correlation
functions at equilibrium \cite{Rammer:1998}. However there is no reason why these relationships
should hold at NE.
Motivated by understanding these NE properties and their use in practical nanoscale
devices, we focus in this paper on a specific physical property: the electronic transport.
In particular, we consider the relationship between the electrical conductance and 
the charge susceptibility in nanoscale junctions. 
We provide a definition for the NE charge susceptibility, which can be measured in experiments, 
and examine in detail its relationship with the full non-linear dynamical conductance.

We show that the \NE charge susceptibility and the dynamical conductance of such a system
are related to each other, though in a different manner than at equilibrium. 
At finite bias, they both contain 
information about the charge fluctuation (induced by the bias) in the electronic resonances. 
However the NE charge susceptibility does \emph{not} contain information about purely inelastic
scattering processes which do not involve charge fluctuations.
By measuring both the conductance and the NE charge susceptibility in the same
experiment, one can identify the nature of scattering processes involved in transport through nanoscale 
junctions. 

We illustrate this property with numerical calculations for a model of a single-molecule
nanojunction in the presence of electron-phonon coupling.
Our results are relevant, but not limited only, for electron-phonon scattering processes.
Other examples could be electron-plasmon, electron-electron, electron-spin excitations scattering
events.
In the following, we first briefly recall the relationship between linear conductance and charge 
susceptibility at equilibrium. Then we derive the corresponding relationship in the NE conditions, 
and present numerical calculations.
 
{\bf Equilibrium response functions:}
Within the linear-response theory of a system at equilibrium \cite{Fetter:1971,Rammer:1998,Lee:2011}, 
the current $I$ is related to a frequency-dependent applied bias $V$
via the linear conductance $g$ as ${I}(\omega)=g(\omega)V(\omega)$.
The linear conductance is a response function obtained from the current-density
correlation function $g(t)=({\rm i}e/\hbar)\expect{[\hat{I}(t),\hat{N}(0)]}\theta(t)$, 
where $\hat{N}$ is the total occupancy operator and $\hat{I}$ is the current operator
$\hat{I}=e{\rm d}{\hat{N}}/{\rm d}t$.
The linear conductance $g$ is directly related to the density-density correlation function
$\chi_c(t)=-{\rm i}\expect{[\hat{N}(t),\hat{N}(0)]}\theta(t)$
by the relation
$g(\omega)={\rm i}\omega \frac{e^2}{\hbar} \chi_c(\omega)$.
$\chi_c$ is also known as the charge susceptibility and
represents the response function of the charge density modifications $\delta n$ due to 
variation of the electrostatic potential $\delta\mathit{v}$: $\delta n = \int \chi_c \delta\mathit{v} $
\cite{Note1}.
In the DC limit, one gets a finite linear conductance $g(\omega\rightarrow 0)$
when the charge susceptibility goes as $\chi_c(\omega)=f(\omega)/\omega$ with $f(0)\ne 0$.
At equilibrium, there is a clear and well-defined relationship between the charge susceptibility
and the linear conductance. However there is no {\it a priori} reason why such a relation should still hold 
at \NE when an applied bias drives the system in a non-linear regime.

{\bf Non-equilibrium charge susceptibility and transport:}
We consider a generic system consisting of a interacting central region $C$, 
the scatterer of interest (e.g. a molecule or a quantum dot), connected to two electrodes,
acting as source and drain. The electrodes are non-interacting Fermi seas at 
their own equilibrium and there is no direct contact between them.
We use \NE \GFs (NEGF) to calculate the electric
current and charge of the system in NE conditions \cite{Meir:1992}.
The system is under a finite, but not small, applied bias and is assumed to
have reached a \NE steady-state which can be described by an effective (pseudo) equilibrium 
steady-state density matrix \cite{Hershfield:1993,Dutt:2011}.

We define the \NE charge susceptibility $\chi_c^{\rm NE}$ in the steady-state as the response 
(not necessarily linear) 
for the modifications of the total electronic occupancy of the central region 
$\delta\expect{n_C}$ due to the changes in the applied bias $\delta V$, i.e. changes in the {\em cause} 
that drives the system out-of-equilibrium \cite{Note2}: 
\begin{equation}
\label{eq:defChiNE}
\chi_c^{\rm NE}(V)=\frac{\partial \expect{n_C^{\rm NE}}}{\partial V} .
\end{equation}
The total occupancy $\expect{n_C^{\rm NE}}$ of the central region $C$ is given by the \NE lesser \GF as
$\expect{n_C^{\rm NE}}=-{\rm i}\int{\rm d}\omega \tr[G^<(\omega)] / 2\pi$, where the trace runs over
the electronic states in the region $C$.

We now examine in detail how $\chi_c^{\rm NE}(V)$
is related to the dynamical conductance $G(V)={\rm d}I/{\rm d}V$. 
The current at the left $L$ interface between the central region $C$ and the $L$
lead is given by the Meir-Wingreen expression \cite{Meir:1992}:

\begin{equation}
\label{eq:IL_MeirWingreen}
\begin{split}
I_L = \frac{{\rm i}  e}{\hbar} \int \frac{{\rm d}\omega}{2\pi}\
{\rm Tr} \left[ f_L(\omega) (G^r_C(\omega) - G^a_C(\omega)) \Gamma_L(\omega) \right. \\
\left. + G^<_C(\omega) \Gamma_L(\omega) \right] ,
\end{split}
\end{equation}
with $\Gamma_L(\omega)/2$ being the imaginary part of the $L$ lead self-energy, and
$G^{r,a,<}_C$ being the retarded, advanced and lesser \GF of the central region
respectively, and the trace is taken over the electron states of the central region.

By using the properties of a NE steady state, one introduces a \NE distribution
functional $f^{\rm NE}_C$ for the central region as 
$G^<_C(\omega)=- f^{\rm NE}_C(\omega) (G^r_C - G^a_C)(\omega)$ \cite{Ness:2010}. At equilibrium
$f^{\rm NE}_C$ is simply given by the conventional Fermi distribution function. 
The dynamical conductance $G(V)$ can be written as:
\begin{equation}
\label{eq:GofV}
\begin{split}
G(V) = \frac{{\rm i}  e}{\hbar} \int \frac{{\rm d}\omega}{2\pi}\
& {\rm Tr} \left[ 
\left(1-f_L (f^{\rm NE}_C)^{-1} \right) \partial_V G^<_C \Gamma_L \right. \\
& \left. - \partial_V \left(f_L (f^{\rm NE}_C)^{-1} \right) G^<_C \Gamma_L \right] ,
\end{split}
\end{equation}
which shows a relation between the dynamical conductance and the derivative of the
lesser \GF versus the applied bias $\partial_V G^<$.
To show more clearly how $G(V)$ and $\chi_c^{\rm NE}(V)$ are 
related to each other, we consider the following simpler system.

{\bf A model system:}
The model consists of a single electron level in the region $C$,
in the presence of some arbritary kind of interaction.
For the moment, we consider the
wideband limit where $\Gamma_L(\omega)=\Gamma$, and that all the potential drop occurs at the left
contact. Only the Fermi distribution $f_L$ of the left lead depends explicitly on the bias 
$V$ via the Fermi level $\mu_L$.
Within these conditions, we find a relation between the dynamical conductance $G$ and the \NE  
charge susceptibility $\chi_c^{\rm NE}$: 

\begin{equation}
\label{eq:GofV_simple1}
\begin{split}
G(V) (\frac{e}{\hbar} \Gamma)^{-1}  + \chi_c^{\rm NE}(V) = \int {\rm d}\omega \partial_V ( f_L A_C(\omega) ) , 
\end{split}
\end{equation}
where 
$A_C(\omega)=(G^a_C(\omega)-G^r_C(\omega))/2\pi{\rm i}$. 

For non-interacting systems, the spectral function $A_C$ is independent of the bias,
then $\partial_V  A_C(\omega)=0$. By using the definitions of 
$G$ and $\chi_c^{\rm NE}$ for symmetric contacts and the corresponding 
\NE distribution function  
$f^{\rm NE}_C = (\Gamma_L f_L +\Gamma_R f_R) /(\Gamma_L + \Gamma_R) = (f_L+f_R)/2$ \cite{Note3},
we find a direct
proportionality between $G$ and $\chi_c^{\rm NE}$: 
$G(V) = \frac{e^2}{\hbar} \Gamma \chi_c^{\rm NE}(V) / e$ \cite{Note4}.
Beyond the wideband approximation (with symmetric contacts), we obtain
the relation: $G(V) = \frac{e^2}{\hbar} \Gamma(\mu_L) \chi_c^{\rm NE}(V) / e$. 
Hence the compatibility between the equilibrium and NE approaches implies that
$\lim_{\omega \rightarrow 0} {\rm i}\omega \chi_c(\omega) \equiv \Gamma \chi_c^{\rm NE}(V)$
(within the DC limit of linear-response).

For interacting systems, $A_C$ depends on $V$ through the interaction self-energy 
$\Sigma_{\rm int}(\omega,V)$. An analytical expression relating $G$ and $\chi_c^{\rm NE}$
is more difficult to obtain \cite{Note5}. However we show next, from numerical calculations
beyond the wideband limit, that
there is a clear relationship between $G(V)$ and $\chi_c^{\rm NE}(V)$ for a model of
interaction self-energy.

{\bf An application:}
For this we have to make a choice for the interactions in the central region $C$. The NE charge 
susceptibility has been briefly studied for a model of electron-electron interaction in
the Anderson impurity model at \NE in \cite{Chao:2011}.
 In the following we consider a model electron-phonon interaction in the central region $C$ 
\cite{Dash:2010,Dash:2011}.
Considering such a model permits us to get several different physical effects:
the renormalization of the electron level but also all the phonon replica (the phonon side-band peaks). 
So effectively we are dealing with a richer model of multi-electronic resonances. 
Such a model includes different inelastic scattering events: those related to 
charge fluctuations in the electronic resonances (resonant elastic and inelastic transmission) and 
those involving off-resonant inelastic scattering by tunneling electrons.
However the relationship derived previously for $G(V)$ and $\chi_c^{\rm NE}(V)$ 
is independent of the nature of the interaction (electron-phonon or electron-electron)
in the central region $C$.

\begin{figure}
  \center
  \includegraphics[clip=,width=\columnwidth]{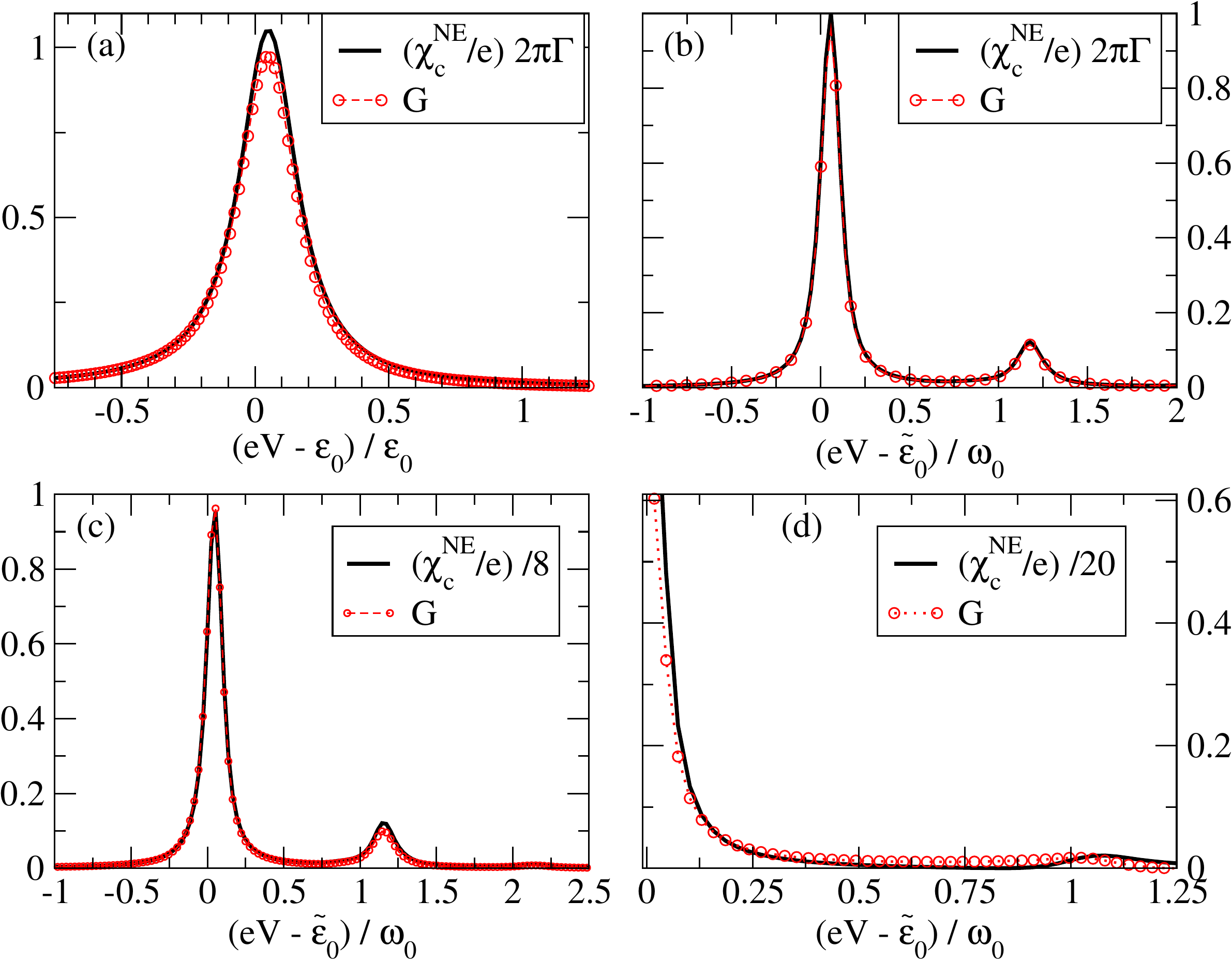}
  \caption{(color online)
           Non-equilibrium charge susceptibility $\chi_c^{\rm NE}$ (full lines)
	   and dynamical conductance $G$ (dashed lines)
	   versus applied bias. 
(a) Non-interaction case, with $\varepsilon_0=0.5$ and $\beta_\alpha=0.7$; 
(b-c) With interaction and for different transport regimes ($\beta_\alpha=2$). 
From off-resonant to resonant:
(b) $\varepsilon_0=0.7$, 
(c) $\varepsilon_0=0.5$, 
(d) $\varepsilon_0=0.15$.
$\chi_c^{\rm NE}$ is rescaled by
$\Gamma=\Gamma_\alpha(\mu^{\rm eq})=t_{0\alpha}^2/\beta_\alpha$.
On this scale both $\chi_c^{\rm NE}$ and $G$ present the same spectral features: peaks associated
with charge fluctuations in the electronic resonances. 
Calculations are done for symmetric coupling  $t_{0\alpha}=0.15$ and asymmetric potential drops
$\mu_L=\mu^{\rm eq}+eV$, $\mu_R=\mu^{\rm eq}$. The other parameters are $\omega_0=0.3$ and
$\gamma_0=0.21$, $\varepsilon_\alpha=0$. The energy parameters are in [eV] and
$G$ in unit of $G_0=e^2/h$.}
  \label{fig:1}
\end{figure}

In our model, the Hamiltonian for the region $C$ is
\begin{equation}
\label{eq:H_central}
\begin{split}
  H_C 
  = \varepsilon_0 d^\dagger d + \omega_0 a^\dagger a +
  \gamma_0 (a^\dagger + a) d^\dagger d,
\end{split}
\end{equation}
where $d^\dagger$ ($d$) creates (annihilates) an
electron in the level $\varepsilon_0$, which
is coupled to  the vibration mode 
of energy $\omega_0$ via the coupling constant $\gamma_0$.
The central region $C$ is connected to two (left and right) one-dimensional
tight-binding chains via the hopping integral $t_{0L}$ and
$t_{0R}$. The corresponding lead $\alpha=L,R$ self-energy is
$\Sigma^r_\alpha(\omega)=t_{0\alpha}^2/\beta_\alpha \exp^{{\rm i} k_\alpha(\omega)}$
with the dispersion relation 
$\omega=\varepsilon_\alpha+2\beta_\alpha \cos(k_\alpha(\omega))$.
Here, the imaginary part $\Gamma_\alpha=-2\Im m \Sigma^r_\alpha$
is energy dependent and goes beyond the wideband limit, unless $\beta_\alpha$ is much
larger than any other paramaters. 
At equilibrium, the whole system has a well-defined unique Fermi level $\mu^{\rm eq}$. 
A finite bias $V$, applied across the junction, lifts the Fermi levels as
$\mu_{L,R}=\mu^{\rm eq}+\eta_{L,R} eV$.  The fraction of potential
drop at the left contact is $\eta_L$ and $\eta_R=\eta_L-1$
at the right contact \cite{Datta:1997}, with $\mu_L-\mu_R=eV$ and 
$\eta_L \in [0,1]$.

Finally the electron-phonon interaction is treated at the Hartree-Fock level
(first order diagrams in term of the interaction) and is incorporated
as self-energies $\Sigma_{\rm eph}^{{\rm HF},r/a/\gtrless}(\omega)$ in the NEGF.
Self-consistent calculations provide a partial resummation of the 
diagrams to all orders \cite{Dash:2010,Dash:2011}.

Within this model, we calculate the dynamical conductance $G(V)$ from 
Eq.~(\ref{eq:IL_MeirWingreen})
and the NE charge susceptibility $\chi_c^{\rm NE}(V)$ from Eq.~(\ref{eq:defChiNE})
for different sets of parameters.
We consider symmetric
($t_{0L}=t_{0R}$) and asymmetric ($t_{0L}\ne t_{0R}$) coupling to the leads, different
strength of coupling to the leads, 
symmetric and asymmetric potential drops at the contacts, 
and
different transport regimes (off-resonant $\varepsilon_0 \ll \mu^{\rm eq}$, and
resonant  $\varepsilon_0 \sim \mu^{\rm eq}$). 
We restrict ourself here to the medium electron-phonon coupling 
($0.5 < \gamma_0/\omega_0 < 1$) regime which corresponds to realistic
coupling in organic molecules. The strong coupling regime requires higher-order
diagrams and more time consuming calculations \cite{Dash:2010,Dash:2011}.

Figure \ref{fig:1} shows the NE charge susceptibility $\chi_c^{\rm NE}(V)$
and the dynamical conductance $G(V)$. We consider
a symmetric coupling to the leads ($t_{0L} = t_{0R}$) and an asymmetric potential drop ($\eta_L=1$).
On this scale, both the conductance and the NE charge susceptibility present 
peaks at an applied bias corresponding to an electronic
resonance: a main resonance peak close to full polaron shift renormalised level 
$\tilde\varepsilon_0=\varepsilon_0-\gamma_0^2/\omega_0$, and phonon sideband peaks
around $V\sim\tilde\varepsilon_0+n\omega_0$ \cite{Note6}.
In the NE conditions, the charge fluctuates
in these electronic resonances whenever the bias window includes $\tilde\varepsilon_0+n\omega_0$.
Hence peaks are obtained in the charge susceptibility $\chi_c^{\rm NE}(V)$ for these biases.
The peaks correspond to elastic
($V\sim\tilde\varepsilon_0$) and inelastic ($V\sim\tilde\varepsilon_0+n\omega_0$) resonant scattering 
processes.
For the non-interacting case Fig.~\ref{fig:1}(a), there is only one resonance at $\varepsilon_0$,
and, as demonstrated, $\chi_c^{\rm NE}$ and $G$ are related via $\Gamma(\mu_L)$ beyond the 
wideband limit.

\begin{figure}
  \center
  \includegraphics[clip=,width=\columnwidth]{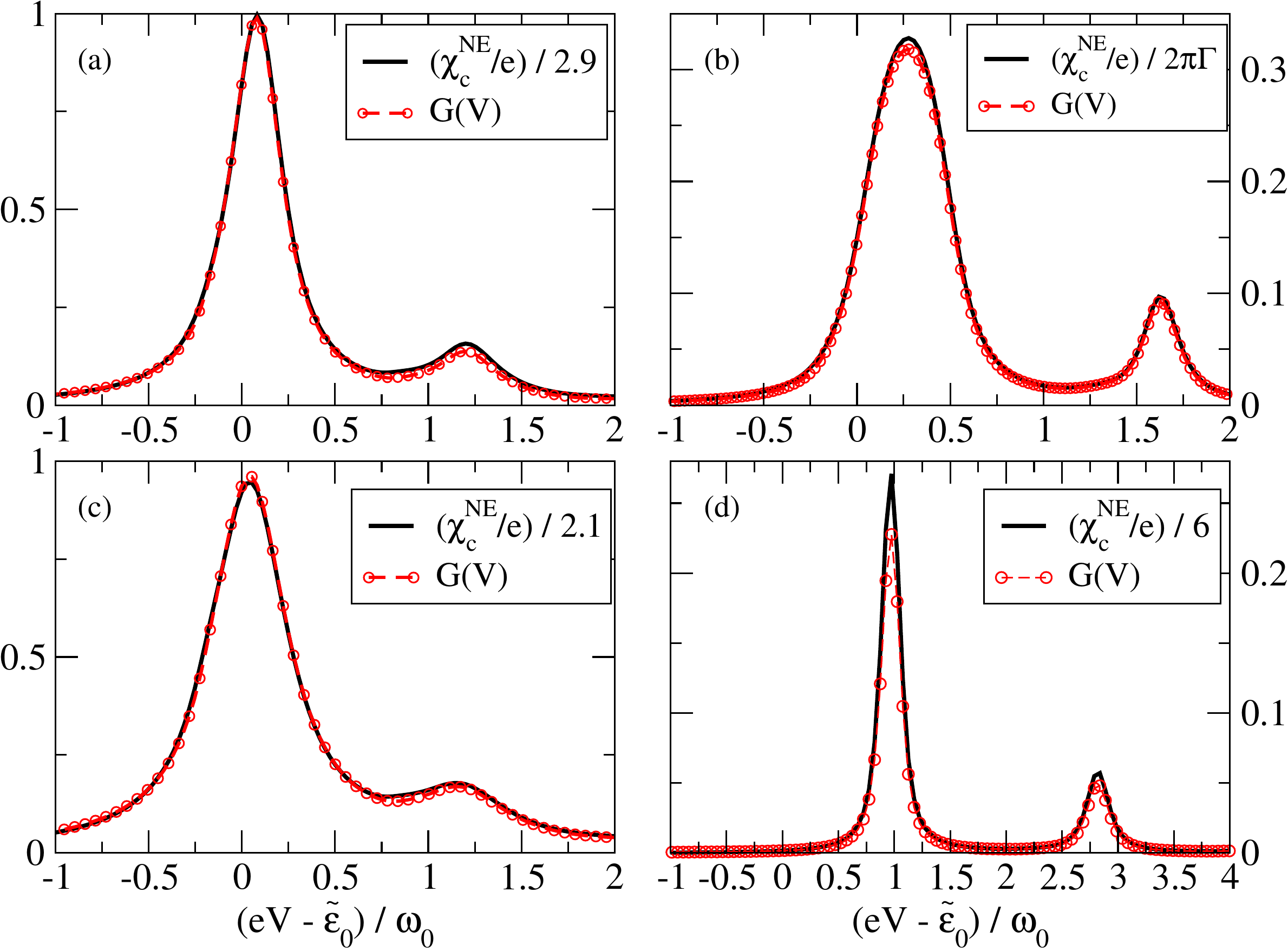}
  \caption{(color online)
           Non-equilibrium charge susceptibility $\chi_c^{\rm NE}$ (full lines)
	   and dynamical conductance $G$ (dashed lines)
	   versus $V$.
Same parameters as in Fig. 1(c) except as otherwise stated.
(a) Far beyond the wideband approximation: $\beta_\alpha=0.7$.
(b) Interaction at the Fock level only.
(c) Strong coupling to the leads $t_{0\alpha}=0.30=\omega_0$.
(d) Asymmetric coupling to the leads and potential drop,
$t_{0L}=0.07, t_{0R}=0.15$ and  
$\eta_L=t_{0R}/(t_{0L}+t_{0R})=0.68182$, and $\varepsilon_0=0.70$.}
  \label{fig:2}
\end{figure}

Figure \ref{fig:2} shows that the relationship between $\chi_c^{\rm NE}$ and $G$
is robust against our model parameters. 
It holds for asymmetric coupling to the leads ($t_{0L}\ne t_{0R}$) and different
fractions of potential drops at the contacts - see Fig 2(d). It holds for strong 
coupling to the leads $t_{0\alpha}\sim\omega_0>\gamma_0$ - see Fig 2(c); and beyond
the wideband limit - see Fig. 2(b). It also holds when the interaction is modelled 
only with the Fock diagram - see Fig. 2(a). Therefore the relation between 
$\chi_c^{\rm NE}$ and $G$ is not due
to the fact that the Hartree self-energy $\Sigma_{\rm eph}^H$ is proportional to
$\expect{n_C^{\rm NE}}$ (hence $\partial_V \Sigma_{\rm eph}^H \propto \chi_c^{\rm NE}$).

Note that with potential drops at both contacts, $\mu_L$ and $\mu_R$ 
support a fraction of the bias, and the relationship between 
$\chi_c^{\rm NE}$ and $G$  includes terms in $\partial_V f_R$.
However both quantities still present the same features versus 
applied bias - see Fig. 2(d).

\begin{figure}
  \center
  \includegraphics[clip,width=\columnwidth]{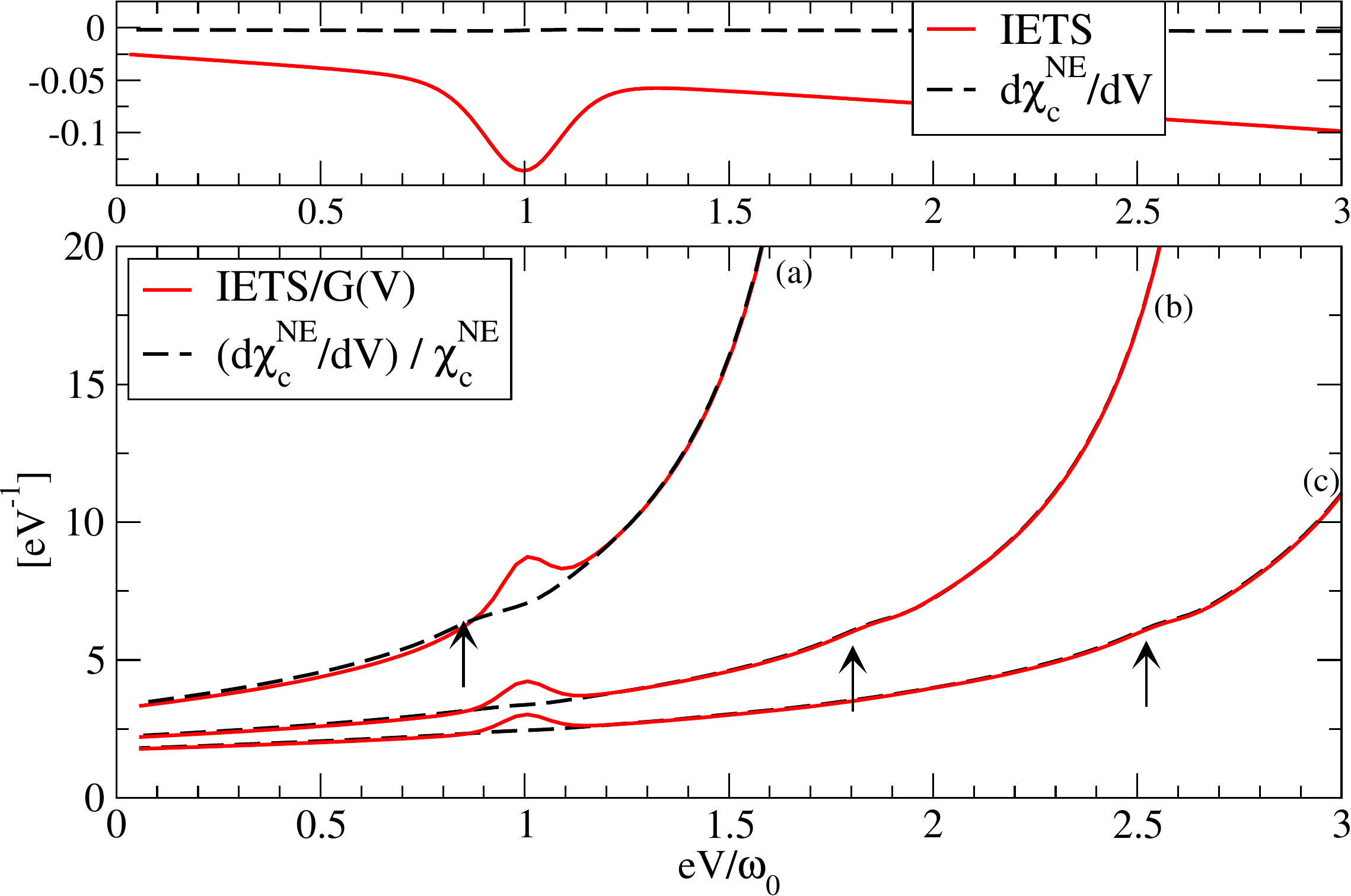}
  \caption{(color online)
           Derivative  $\partial_V\chi_c^{\rm NE}$ (dashed lines)
	   and IETS signal $\partial_V G$ (full lines).
	   $\partial_V\chi_c^{\rm NE}$ does not have the peak or dip
	   feature of the IETS at $V\sim\omega_0$.
(\emph{Top})
Resonant transport regime. Calculations done with
$t_{0\alpha}=1.50\sim\beta_\alpha$, $\varepsilon_0=0.0$, $\gamma_0=0.195$.
(\emph{Bottom})
Off-resonant regime for different $\varepsilon_0$.
$\partial_V\chi_c^{\rm NE}$ and $\partial_V G$
are normalised by $\chi_c^{\rm NE}$ and $G$ respectively.
(a) $\varepsilon_0=0.70$, (b) $\varepsilon_0=0.99$, (c) $\varepsilon_0=1.20$.
The arrows point the position of the electron resonance at 
$V=\tilde\varepsilon_0-\omega_0$. Calculations done with the same parameters as in Fig. 1(b).}
  \label{fig:3}
\end{figure}

On a smaller energy scale, the conductance also contains physical information for biases around 
excitation energies which goes beyond resonant transmission. 
Indeed, the conductance also varies at bias thresholds corresponding to other inelastic scattering
processes (for example inelastic electron tunneling).
At the bias threshold $V\sim\omega_0$, the conductance increases in the off-resonant transport regime
(opening of new conduction channels) or decreases in the resonant transport regime (electron-phonon
backscattering).
These effects are better seen in the inelastic electron tunneling spectroscopy (IETS) as peaks 
or dips for the off-resonance or resonant transport regime respectively \cite{Galperin:2007b,Dash:2011}. 
The IETS is obtained from the second derivative of the
current versus applied bias ${\rm d}^2I/{\rm d}V^2={\rm d}G/{\rm d}V\equiv\partial_V G(V)$. 
In experiments, the IETS signal is usually given normalised by the conductance
itself or by the current itself.
Figure \ref{fig:3} shows the IETS signal as well as the corresponding
variation of the \NE charge susceptibility versus applied bias $\partial_V \chi_c^{\rm NE}$.
One can clearly see a peak feature at $V\sim\omega_0$ in the IETS signal, while $\chi_c^{\rm NE}$ is 
virtually featureless at the corresponding bias for both the off-resonant and resonant transport regimes. 
This means that these inelastic tunneling electron-phonon scattering processes (at
$V\sim\omega_0$) are not related to charge fluctuations.
Instead the phonon population fluctuates because of phonon emission induced by the tunneling electron.
Note that the tiny features pointed by the arrows in Figure \ref{fig:3} correspond to tiny peak 
features in both $\chi_c^{\rm NE}$ and $G$. They are related to charge fluctuations in the
electron resonances at $V=\tilde\varepsilon_0-\omega_0$ (phonon emission by a hole).

{\bf Discussion: }
We have hence shown that the non-equilibrium charge susceptibility and the 
dynamical conductance are directly related to each other, though in a different manner
than for the equilibrium case. 
They both present features (peaks) versus the applied bias whenever
there are charge fluctuations in the corresponding electronic resonances of the nanojunction.

Therefore we suggest that measuring both the conductance and the NE charge 
susceptibility simultaneously in the same experiment is essential in quantum transport. 
It permits one to identify the nature of the scattering 
processes involve in the transport, i.e. processes involving charge fluctuation
or not.
This result is very important for the analysis of the transport properties in complex
systems such as large single-molecule junction and does not involve the presence of a third
gate electrode.
Although our result is mostly relevant for electron-phonon scattering processes, it is
not limited only to these processes. 
The measurement of the NE charge susceptibility could be performed by measuring the potential drop around 
a capacitor placed in series with the nanojunctions ($V_{\rm cap} = e\expect{n_C}/C_{\rm cap}$). 
One can then obtain $\chi_c^{\rm NE}(V)$ in a similar way as the dynamical 
conductance $G(V)$ is obtained from the current by using a lock-in set up.

\end{document}